# Effects of localized source currents on magnetotelluric responses of a flat earth and spherical earth

Shinya Sato

Graduate School of Engineering, Kyoto University

Email: sato.shinya.27s@st.kyoto-u.ac.jp

**Abstract**

Shifts in magnetotelluric (MT) responses owing to localized source currents should be considered when visualizing deep subsurface resistivity structures such as the earth's mantle. The earth is not flat but spherical; nevertheless, non-uniform geomagnetic temporal variations arising from localized source currents are evaluated on the basis of components in a Cartesian coordinate system. To address this issue, this study assesses the difference in source bias within MT responses of a flat earth and spherical earth. Apparent resistivity and phases are calculated by setting the time period and conductivity of the earth's interior to 200/2000 s and 0.001 S/m, respectively, and by changing the vertical and horizontal distances between the source current and an observation station. A deviation in the biased MT responses of the flat earth and spherical earth is not observed although both shift strongly from the true values. We can thus treat the source bias in a Cartesian coordinate system although it originally arises in a spherical coordinate system.

**Introduction**

In magnetotelluric (MT) surveys, the primary electromagnetic (EM) fields arising from source currents are assumed to be horizontally uniform. Many previous studies (Madden and Nelson 1964; Schmucker 1970; Hermance and Peltier 1970; Häkkinen et al. 1989; Pirjola 1992; Viljanen 2012) have reported that the MT responses at long periods and sites above resistive structures are biased owing to localized source currents. These studies focused on period-dependent bias, whereas Sato (2020a) and Sato (2020b) calculated the bias by varying the horizontal/vertical distances of the localized source currents from an observation station in detail. Because the distances can vary temporally (Sato 2020a, 2020b, and references therein), we should consider the source-dependent shifts in MT responses when visualizing deep and resistive subsurface structures like the earth's mantle.

Localized source currents are expected to cause horizontally non-uniform geomagnetic temporal variations. Murphy and Egbert (2018) and Wang et al. (2020) showed the relationship between the biased MT responses and spatial gradients of geomagnetic fields using the method introduced by Egbert (1999). Sato et al. (2020) presented a new method to directly assess the spatial gradients from raw geomagnetic spectrograms using non-negative matrix factorization (Kameoka et al. 2009; Kitamura 2019). These techniques are applied to $x/y$ components in a Cartesian coordinate system. However, one could argue that it would be better to evaluate the spatial gradients in a spherical coordinate system (i.e., $\theta/\varphi$ components) because the earth is not flat but spherical. Thus, this study tests the difference of shifts in MT responses of a flat earth and spherical earth owing to the localized source currents.

The study presents: (i) numerical examples of biased MT responses of a flat earth and spherical earth by changing the vertical/horizontal distances between a site and the source current; (ii) interpretations from these examples; and (iii) the mathematical background responsible for the difference of shifts in the MT impedances of both earth types.

**EM fields on the earth's surface**

We use a line current and loop current in the flat earth for an equatorial electrojet (EEJ) and Sq current, respectively. Additionally, we assume four types of source current and earth: (a) EEJ and flat earth; (b) Sq current and flat earth; (c) EEJ and spherical earth; and (d) Sq current and spherical earth. The coordinate systems and source currents used for those cases are depicted in Fig. 1. In the coordinate systems for cases (a) and (b), the $x$-, $y$-, and $z$-axes are northward, westward, and downward positive, respectively, with $z = 0$ at the earth's surface. In cases (c) and (d), the earth's surface is defined as $r = a$. The $y$-axis corresponds to the north pole. Because the Sq current does not flow around the north pole, the coordinate system in case (d) is modified, as described later.

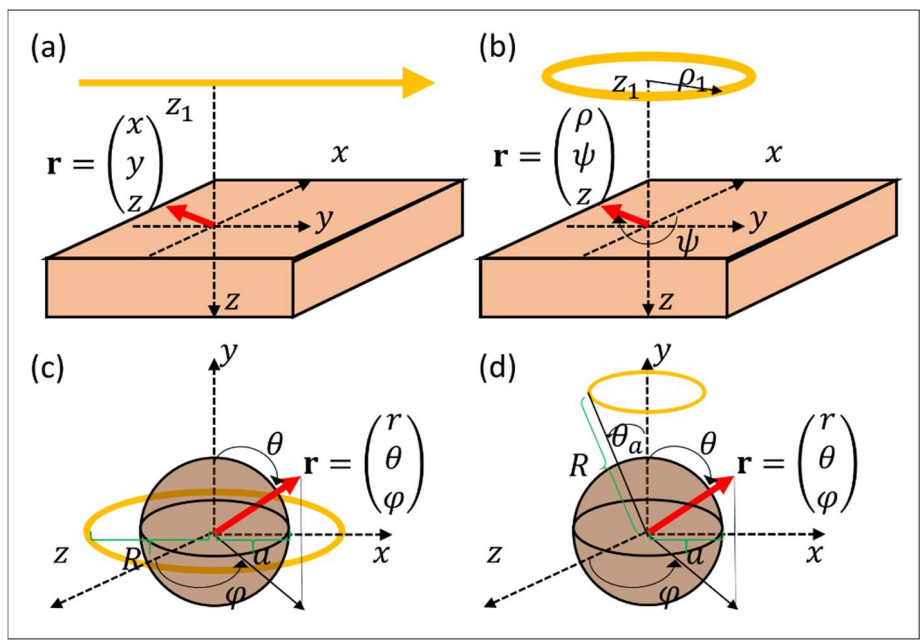

Figure 1. Coordinate systems used for cases (a)–(d). **r** denotes a position vector.

Let us consider an example using the SI system. The source currents used for cases (a), (b),

and (d) are defined as:

$$\mathbf{J}^o = \begin{pmatrix} J_x^o \\ J_y^o \\ J_z^o \end{pmatrix} = I\delta(x)\delta(z-z_1)\begin{pmatrix} 0 \\ 1 \\ 0 \end{pmatrix}, \tag{1}$$

$$\mathbf{J}^o = \begin{pmatrix} J_\rho^o \\ J_\psi^o \\ J_z^o \end{pmatrix} = I\delta(\rho-\rho_1)\delta(z-z_1)\begin{pmatrix} 0 \\ 1 \\ 0 \end{pmatrix}, \tag{2}$$

$$\mathbf{J}^o = \begin{pmatrix} J_r^o \\ J_\theta^o \\ J_\varphi^o \end{pmatrix} = I\delta(r-R)\delta(\theta-\theta_a)\begin{pmatrix} 0 \\ 0 \\ 1 \end{pmatrix}, \tag{3}$$

respectively, where $I$ is an electric current. The current density for case (c) is obtained by setting $\theta_a$ in Eq. (3) to $\frac{\pi}{2}$. Let us consider a flat earth structure with a half-space of conductivity $\sigma_1$, and that of the spherical earth with a sphere of radius $a$ and conductivity $\sigma_1$. The EM fields on the earth's surface and apparent resistivity can be represented at $(x, y, 0)$ for case (a) as:

$$E_y = -i\omega\frac{\mu_0 I}{2\pi}\int_{-\infty}^{\infty}\frac{1}{|\eta|+\sqrt{\eta^2+i\omega\mu_0\sigma_1}}e^{|\eta|z_1}e^{-i\eta x}d\eta, \tag{4}$$

$$B_x = \frac{\mu_0 I}{2\pi}\int_{-\infty}^{\infty}\frac{\sqrt{\eta^2+i\omega\mu_0\sigma_1}}{|\eta|+\sqrt{\eta^2+i\omega\mu_0\sigma_1}}e^{|\eta|z_1}e^{-i\eta x}d\eta, \tag{5}$$

$$\rho_{xy} = \frac{\mu_0}{\omega}\left|\frac{E_y}{B_x}\right|^2, \tag{6}$$

those at $(\rho, \psi, 0)$ for case (b) as:

$$E_\psi = -i\omega\mu_0 I\rho_1 \int_0^\infty \frac{\eta}{\eta+\sqrt{\eta^2+i\omega\mu_0\sigma_1}}e^{\eta z_1}J_0(\eta\rho)J_0(\eta\rho_1)d\eta, \tag{7}$$

$$B_\rho = \mu_0 I\rho_1 \int_0^\infty \frac{\eta\sqrt{\eta^2+i\omega\mu_0\sigma_1}}{\eta+\sqrt{\eta^2+i\omega\mu_0\sigma_1}}e^{\eta z_1}J_0(\eta\rho)J_0(\eta\rho_1)d\eta, \tag{8}$$

$$\rho_{\psi\rho} = \frac{\mu_0}{\omega}\left|\frac{E_\psi}{B_\rho}\right|^2, \tag{9}$$

and those at $(a, \theta, \varphi)$ for cases (c) and (d) as:

$$E_\varphi = i\omega\frac{\mu_0 I}{2}\sin\theta_a \sum_{n=0}^{\infty}(2n+1)P_n(\cos\theta_a)P_n(\cos\theta)\left\{\frac{1}{k}\cdot\frac{J_{n+\frac{1}{2}}(ka)}{J_{n-\frac{1}{2}}(ka)}\right\}\frac{a^{n-1}}{R^{n-1}}, \tag{10}$$

$$B_\theta = \frac{\mu_0 I}{2}\sin\theta_a \sum_{n=0}^{\infty}(2n+1)P_n(\cos\theta_a)P_n(\cos\theta)\left\{1-\frac{n}{ka}\cdot\frac{J_{n+\frac{1}{2}}(ka)}{J_{n-\frac{1}{2}}(ka)}\right\}\frac{a^{n-1}}{R^{n-1}}, \tag{11}$$

$$\rho_{\varphi\theta} = \frac{\mu_0}{\omega}\left|\frac{E_\varphi}{B_\theta}\right|^2, \tag{12}$$

where $\mu_0$, $\omega$, $J_\nu$, $P_n$, and $k$ are the magnetic permeability of free space, angular frequency, Bessel function of the first kind for order $\nu \in \mathbb{R}$, Legendre polynomials for order $n$, and equal to $\sqrt{-i\omega\mu_0\sigma_1}$, respectively. Eqs. (4) and (5) and Eqs. (10) and (11) are derived following Sato (2020a) and Sato (2020b), respectively. Eqs. (7) and (8) are obtained as shown in Appendix A.

Hereafter, this paper represents parameters $x$, $\rho$, and $\rho_1$ in cases (a) and (b) in degree, which is set to 110 km to facilitate a comparison with the results from cases (c) and (d). The variables $\psi$, $\theta$, $\theta_a$, and $\varphi$ are also unified in degree. The altitudes of the source currents are defined as the distance from the earth's surface and represented in km instead of m for simplicity. When calculating the MT impedances, we may neglect $y$ in case (a), $\psi$ in case (b), $\varphi$ in cases (c) and (d), and $I$ in all cases. Additionally, the distance of the source current from an observation station is shortened to "the distance".

**MT responses at the earth's surface**
The conductivity of the earth $\sigma_1$ and time period $T$ are set to 0.001 S/m and 200 s, respectively. The subsurface conductivity has the same value as that used for the crust in Hermance and Peltier (1970). Because the mantle conductivity and period used to visualize the mantle are 0.001–0.0001 S/m and 100–10000 s, respectively, the above values are reasonable. The altitudes of the source current in cases (a)–(d) are varied from 100 to 600 km in 10-km increments. This range corresponds to the E (100–150 km) and F regions (150–600 km). In cases (c) and (d), the earth's radius $a$ is set to 6,400 km.

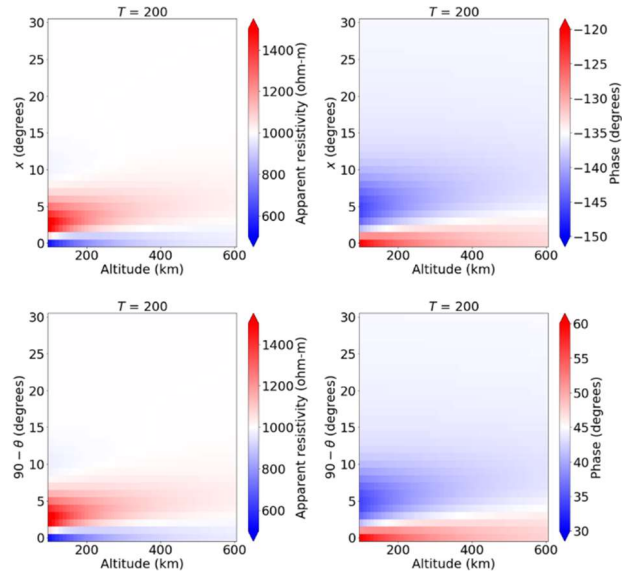

Figure 2. Upper and lower panels denote MT responses of the flat earth and spherical earth, respectively, when using an EEJ for the source current. Left and right panels are apparent resistivity and phase, respectively.

Here, we compare cases (a) and (c) where an EEJ is used for the source current. The parameters $x$ and $\theta$ are varied from 0° to 30° and from 90° to 60° (i.e., from 0° to 30° N latitude), respectively. The integrals in Eqs. (4) and (5) are calculated using the discrete approximation and the convergence of each is checked. The summations in Eqs. (10) and (11) are calculated up to $n = 1000$. The results are shown in Fig. 2. The MT responses of both earth type shifts strongly depend on the horizontal/vertical distance, whereas such bias appears equal.

We compare cases (b) and (d), in which Sq current is used for the source current. The observation station is assumed to be Japan that is located at 35° N. The focus latitude of the Sq current varies from 20° to 45° N (Yamazaki and Maute 2017). For calculating the EM fields under these conditions, we vary $\rho$ from –10° to 15° in case (b) because the source's center corresponds to the $z$-axis, as shown in Fig. 1b. In case (d), the coordinate system is rotated so that the source's focus corresponds to the $y$-axis, as shown in Fig. 1d. As a result, we change $\theta$ from –10° to 15°, as case (b). In both cases, the variation range is set to 1°–15° because of the axial symmetry. The Sq current flow is high especially in the range within the circle with a radius of 10°–15° (Yamazaki and Maute 2017). We therefore set $\rho_1$ and $\theta_a$ to 10°. The integrals in Eqs. (7) and (8) are calculated using the discrete approximation by checking their convergences. The results are shown in Fig. 3. The biases within the MT responses of the flat and spherical earth models due to the Sq current are large but equal.

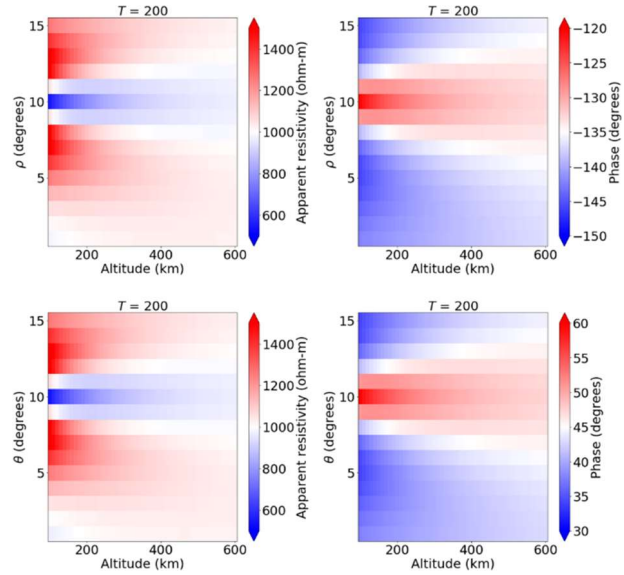

Figure 3. Same as in Fig. 2 but derived using the Sq current for the source current.

Three features can be observed in the above results (Figs. 2, 3) as: (1) the apparent resistivity and phase shift strongly from the true values and depend on the vertical/horizontal distance; (2) bias decreases with increasing distance; and (3) the shifts in the MT responses of the flat earth are the same as those of the spherical earth. These features are also observed in the MT impedances at the time period of 2,000 s (not shown).

**Discussion**

Here we discuss the interpretation of the above results (Figs. 2, 3) and the condition that the bias within the MT responses of the spherical earth deviates from those of the flat earth. The reason for the second feature mentioned above is well explained in Sato (2020a), although limited to case (a).

The skin depth, where the EM fields are attenuated by 1/e, is defined as $503\sqrt{\frac{T}{\sigma_1}}$ (Chave and Jones 2012). Substituting 0.001 S/m and 200/2000 s into $\sigma_1$ and $T$, respectively, yields a skin depth of over 200/2000 km, which corresponds to the asthenosphere/mesosphere. We must thus consider the bias within MT responses that arise from the localized source currents (Figs. 2, 3) if the target to be visualized is the earth's mantle. As introduced above, some methods (Egbert 1999; Sato et al. 2020) can be used to evaluate the spatial gradients of geomagnetic fields arising from localized source currents. These methods are applied to the *x*/*y* components of the MT/geomagnetic data (i.e., in a Cartesian coordinate system). Because of the equality of the source bias within the MT responses of both earth types (Figs. 2, 3), the above methods are considered effective for detecting shifts owing to localized source currents.

The time period is set to 20,000 s to assess the difference of bias within the MT responses of the flat earth and spherical earth although in actuality we cannot assume that the distance is constant because of the earth's rotation. For simplicity, we focus only on the results obtained from cases (a) and (c). The results are shown in Fig. 4 where contour diagrams of apparent resistivity are omitted.

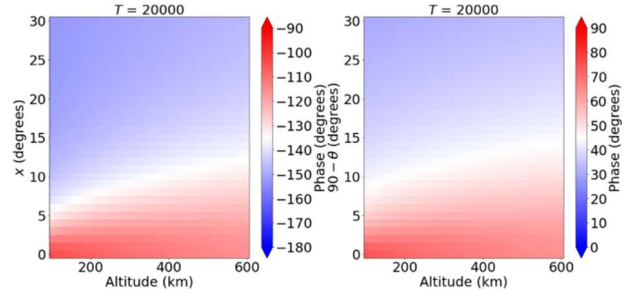

Figure 4. Left: flat earth phase. Right: spherical earth phase. An EEJ is used for the source current.

The deviation of the MT response shifts of both earth types is notable. Such features are also found in the biased apparent resistivity (not shown). The mathematical background regarding this feature is explained by focusing on the phase. As explained in Sato (2020a), by integrating by parts, the MT impedance of the flat earth derived from a line source current (Fig. 1a) can be represented by:

$$Z_{yx} = -i\omega \frac{\left\{\frac{1}{k'(z_1-ix)}+\frac{1}{k'^2(z_1-ix)^2}+\frac{1+k'^3 D_1}{k'^3(z_1-ix)^3}+\frac{1}{k'(z_1+ix)}+\frac{1}{k'^2(z_1+ix)^2}+\frac{1+k'^3 D_2}{k'^3(z_1+ix)^3}\right\}}{\left\{\frac{1}{(z_1-ix)}+\frac{1}{k'(z_1-ix)^2}+\frac{2+k'^2 C_1}{k'^2(z_1-ix)^3}+\frac{1}{(z_1+ix)}+\frac{1}{k'(z_1+ix)^2}+\frac{2+k'^2 C_2}{k'^2(z_1+ix)^3}\right\}}, \quad (13)$$

where $k' = \sqrt{i\omega\mu_0\sigma_1}$, and $C_1$, $C_2$, $D_1$, and $D_2$ are constant values. The observation station is located at $(x, y, 0)$. In the limit $|x| \to \infty$ or $z_1 \to -\infty$, the plane-wave assumption holds:

$$Z_{yx} = \frac{-\sqrt{i\omega}}{\sqrt{\mu_0\sigma_1}}. \quad (14)$$

As a result, if the plane-wave assumption is established, the phase becomes 135°, independent of $\omega$ (i.e., $T$) and $\sigma_1$.

The MT impedances derived from case (c) (Fig. 1c) are:

$$Z_{\varphi\theta} = i\omega \frac{\sum_{n=0}^{\infty}(2n+1)P_n(\cos\theta_a)P_n(\cos\theta)\left\{\frac{1}{ka}\frac{J_{n+\frac{1}{2}}(ka)}{J_{n-\frac{1}{2}}(ka)}\right\}\frac{a^{n-2}}{R^{n-1}}}{\sum_{n=0}^{\infty}(2n+1)P_n(\cos\theta_a)P_n(\cos\theta)\left\{1-\frac{n}{ka}\frac{J_{n+\frac{1}{2}}(ka)}{J_{n-\frac{1}{2}}(ka)}\right\}\frac{a^{n-1}}{R^{n-1}}}. \tag{15}$$

We replace $ka$ by $\lambda$ and introduce $C_n(\lambda)$:

$$C_n(\lambda) = \frac{1}{\lambda} \cdot \frac{J_{n+\frac{1}{2}}(\lambda)}{J_{n-\frac{1}{2}}(\lambda)}. \tag{16}$$

When $n = 0$, $C_0(\lambda)$ is equal to $\frac{1}{\lambda}\tan\lambda$. Let the time period approach infinity as $T \to \infty$. Because $\lambda$ is represented by $\sqrt{-i\frac{2\pi}{T}\mu_0\sigma_1}a$, we may consider that it becomes zero. As a result, we obtain:

$$\lim_{\lambda\to 0} C_0(\lambda) = \lim_{\lambda\to 0}\frac{1}{\lambda}\tan\lambda = 1. \tag{17}$$

When $n = 1, 2, ...$, we represent $J_{n-\frac{1}{2}}(\lambda)$ by:

$$J_{n-\frac{1}{2}}(\lambda) = \frac{\left(\frac{1}{2}\lambda\right)^{n-\frac{1}{2}}}{\pi^{\frac{1}{2}}\Gamma(n)}\int_{-1}^{1}(1-t^2)^{n-1}e^{i\lambda t}dt, \tag{18}$$

where $\Gamma$ denotes a gamma function that holds $n\Gamma(n) = \Gamma(n+1)$ $(n = 1,2,...)$. Using a real number $p$, $\lambda$ is written as:

$$\lambda = (1-i)p. \tag{19}$$

By substituting Eqs. (18) and (19) into Eq. (16), $C_n(\lambda)$ is given as:

$$C_n(\lambda) = \frac{1}{2n} \cdot \frac{\int_{-1}^{1}(1-t^2)^n e^{ipt}e^{pt}dt}{\int_{-1}^{1}(1-t^2)^{n-1}e^{ipt}e^{pt}dt}. \tag{20}$$

We may consider the limit $p \to 0$ instead of $\lambda \to 0$ because of Eq. (19). Let a function $F_{n,p}(t)$ defined on $-1 \leq t \leq 1$ satisfy:

$$F_{n,p}(t) = (1-t^2)^n e^{ipt}e^{pt}. \tag{21}$$

The interval on $p$ may be defined as $(0,1]$ because of the limit $p \to 0$, in which $F_{n,p}(t)$ convergences pointwise to $F_n(t) = (1-t^2)^n$. Additionally, within the range of $[-1,1]$, the relationship:

$$\left|F_{n,p}(t)\right| = \left|(1-t^2)^n e^{ipt}e^{pt}\right| \leq e, \tag{22}$$

is established. Thus, on the basis of Lebesgue's dominated convergence theorem, we can obtain:

$$\lim_{p\to 0}\frac{1}{2n} \cdot \frac{\int_{-1}^{1}(1-t^2)^n e^{ipt}e^{pt}dt}{\int_{-1}^{1}(1-t^2)^{n-1}e^{ipt}e^{pt}dt} = \frac{1}{2n} \cdot \frac{\int_{-1}^{1}(1-t^2)^n dt}{\int_{-1}^{1}(1-t^2)^{n-1}dt}. \tag{23}$$

The right-hand side of Eq. (23) is a "real" constant value.

In the limit $T \to \infty$, the numerator and denominator of the right-hand side of Eq. (15) are

"real" because of Eqs. (17) and (23). As a result, the phase of the spherical earth has a constant value of 90°. Although the time period used for Fig. 4 is not particularly long, the cause of the derivation in the phases of the flat earth and spherical earth can be considered as triggered, as described above.

**Summary**

We calculated EM fields by assuming four cases of source current type and earth model: (a) EEJ and flat earth; (b) Sq current and flat earth; (c) EEJ and spherical earth; and (d) Sq current and spherical earth. The MT response shift in all cases strongly depends on the horizontal or vertical distance between the source current and site. We should consider such source bias when visualizing deep subsurface structures like the earth's mantle. However, the shifts in the MT responses of the flat earth and spherical earth are equal. Thus, we can treat the source bias in a Cartesian coordinate system although the actual earth is not flat but spherical. As a result, unless the target period is extremely long, the conventional or recent-proposed methods to evaluate the spatial gradients of geomagnetic fields from *x*/*y* components (i.e., Cartesian coordinate system) can be considered effective for detecting the source bias.

**Appendix A**

Here we derive the EM fields in case (b). Because we can ignore the displacement current in the MT method, Maxwell's equations in the frequency domain are:

$$\nabla \times \mathbf{E} = -i\omega \mathbf{B}, \tag{A1}$$

$$\nabla \times \mathbf{B} = \mu_0(\sigma \mathbf{E} + \mathbf{J}^o), \tag{A2}$$

$$\nabla \cdot \mathbf{B} = 0, \tag{A3}$$

where $\sigma$ denotes conductivity. We introduce the vector potential $\mathbf{A}$ and scalar potential $\Pi$, and the EM fields are represented by:

$$\mathbf{B} = \nabla \times \mathbf{A}, \tag{A4}$$

$$\mathbf{E} = -i\omega(\mathbf{A} + \nabla \Pi). \tag{A5}$$

By applying a gauge transformation such that $\mathbf{A}$ and $\Pi$ satisfy:

$$i\omega\sigma\mu_0 \Pi = -\nabla \cdot \mathbf{A}, \tag{A6}$$

the equation for $\mathbf{A}$ is given by:

$$-\Delta \mathbf{A} + i\omega\sigma\mu_0 \mathbf{A} = \mu_0 \mathbf{J}^o. \tag{A7}$$

Additionally, because of $\nabla \cdot \mathbf{E} = 0$, we can obtain the equation for the scalar potential as:

$$-\Delta \Pi + i\omega\sigma\mu_0 \Pi = 0. \tag{A8}$$

Let us consider the EM fields above the earth's surface. Because $\sigma$ may be taken as zero, by substituting Eq. (2) into $\mathbf{J}^o$, Eq. (A7) is given as:

$$-\Delta A_\psi = \mu_0 J_\psi^o = \mu_0 I \delta(\rho - \rho_1)\delta(z - z_1). \tag{A9}$$

Because $A_\psi$ is independent of $\psi$, both sides of Eq. (A6) vanish. We can thus neglect the scalar potential.

The Fourier transform (FT) for horizontal components $\rho$ and $\psi$ and its inverse (IFT) are defined in this study as:

$$\tilde{f}(\eta,\varphi) = \int_0^\infty \int_0^{2\pi} \rho f(\rho,\psi) e^{-i\eta\rho\cos(\varphi-\psi)} d\psi d\rho, \tag{A10}$$

$$f(\rho,\psi) = \left(\frac{1}{2\pi}\right)^2 \int_0^\infty \int_0^{2\pi} \eta \tilde{f}(\eta,\varphi) e^{i\eta\rho\cos(\varphi-\psi)} d\varphi d\eta. \tag{A11}$$

The application of the FT in Eq. (A10) to Eq. (A9) yields:

$$\frac{\partial^2}{\partial z^2}\widetilde{A_\psi}(\eta,\varphi,z) - \eta^2 \widetilde{A_\psi}(\eta,\varphi,z) = -\mu_0 \widetilde{J_\psi^o}(\eta,\varphi,z), \tag{A12}$$

where $\widetilde{A_\psi}$ and $\widetilde{J_\psi^o}$ are $A_\psi$ and $J_\psi^o$ are in the wave number domain. Let us consider the Green function for Eq. (A12), which satisfies:

$$\frac{\partial^2}{\partial z^2} G(z,z') - \eta^2 G(z,z') = \delta(z-z'). \tag{A13}$$

As shown in Arfken et al. (2012), the solution to Eq. (A13) is:

$$G(z,z') = -\frac{e^{\eta(z'-z)} + \Omega e^{\eta(z'+z)}}{2\eta} \quad (0 \geq z \geq z'), \tag{A14}$$

where $\Omega$ is a constant upholding the boundary condition at $z=0$. As a result, the solution for $\widetilde{A_\psi}$ can be written as:

$$\widetilde{A_\psi}(\eta,\varphi,z) = \int_{-\infty}^0 \frac{\mu_0}{2\eta}\{e^{\eta(z'-z)} + \Omega e^{\eta(z'+z)}\}\widetilde{J_\psi^o}(\eta,\varphi,z')dz'. \tag{A15}$$

Using the FT in Eq. (A10) for $J_\psi^o$, $\widetilde{J_\psi^o}(\eta,\varphi,z)$ is given by:

$$\widetilde{J_\psi^o}(\eta,\varphi,z) = I\rho_1 \delta(z-z_1) \int_0^{2\pi} e^{-i\eta\rho_1 \cos(\varphi-\psi)} d\psi. \tag{A16}$$

Because the integrals in Eq. (A16) can be replaced by a Bessel function of the first kind $2\pi J_0(\eta\rho_1)$, we obtain:

$$\widetilde{J_\psi^o}(\eta,\varphi,z) = 2\pi I\rho_1 \delta(z-z_1) J_0(\eta\rho_1). \tag{A17}$$

Substituting Eq. (A17) into Eq. (A15), $\widetilde{A_\psi}(\eta,\varphi,z)$ is given by:

$$\widetilde{A_\psi}(\eta,\varphi,z) = \mu_0 \pi I \rho_1 \frac{e^{\eta(z_1-z)} + \Omega e^{\eta(z_1+z)}}{\eta} J_0(\eta\rho_1). \tag{A18}$$

We apply the IFT in Eq. (A11) to Eq. (A18) and obtain:

$$A_\psi(\rho,\psi,z) = \frac{\mu_0 I \rho_1}{2} \int_0^\infty \{e^{\eta(z_1-z)} + \Omega e^{\eta(z_1+z)}\} J_0(\eta\rho) J_0(\eta\rho_1) d\eta. \tag{A19}$$

Because $B_\rho = -\frac{\partial}{\partial z} A_\psi$ and $E_\psi = -i\omega A_\psi$, the EM fields are written as:

$$E_\psi = -i\omega \frac{\mu_0 I \rho_1}{2} \int_0^\infty \{e^{\eta(z_1-z)} + \Omega e^{\eta(z_1+z)}\} J_0(\eta\rho) J_0(\eta\rho_1) d\eta, \tag{A20}$$

$$B_\rho = -\frac{\mu_0 I \rho_1}{2} \int_0^\infty \eta\{-e^{\eta(z_1-z)} + \Omega e^{\eta(z_1+z)}\} J_0(\eta\rho) J_0(\eta\rho_1) d\eta. \tag{A21}$$

Let us consider the earth's interior having a half-space of conductivity $\sigma_1$. The continuity of the EM fields yields the relationship:

$$\Omega = \frac{\eta - \sqrt{\eta^2 + i\omega\sigma_1\mu_0}}{\eta + \sqrt{\eta^2 + i\omega\sigma_1\mu_0}}. \tag{A22}$$

From Eq. (A20)–(A22), we can obtain the EM fields (Eqs. (7), (8)) arising from a loop current (Fig. 1b).

**Reference**

Arfken GB, Weber HJ, Harris FE (2012) Mathematical methods for physicists: A comprehensive guide. Academic Press, Cambridge.

Chave AD, Jones AG (2012) Introduction to the magnetotelluric method. In: Chave AD, Jones AG (ed) The Magnetotelluric Method. Cambridge University Press, Cambridge.

Egbert GD (1997) Robust multiple-station magnetotelluric data processing. Geophys J Int 130(2): 475–496.

Häkkinen L, Pirjola R, Sucksdorff C (1989) EISCAT magnetometer cross and theoretical studies connected with the electrojet current system. Geophysica 25(1): 123–134.

Hermance JF, Peltier WR (1970) Magnetotelluric fields of a line current. J Geophys Res 75(17): 3351–3356.

Kameoka H, Ono N, Kashino K, Sagayama S (2009) Complex NMF: A new sparse representation for acoustic signals. In: Acoustics, Speech and Signal Processing, 2009. ICASSP 2009. IEEE International Conference, Taipei, 3437–3440 April 2009.

Kitamura D (2019) Nonnegative matrix factorization based on complex generative model. Acoust. Sci. Technol 40(3): 155–161.

Madden T, Nelson P (1964) A defense of Cagniard's magnetotelluric method. ONR Rept: 371–401.

Murphy BS, Egbert GD (2018) Source biases in midlatitude magnetotelluric transfer functions due to Pc3-4 geomagnetic pulsations. Earth Planets Space 70(1), doi.org/10.1186/s40623-018-0781-0.

Pirjola R (1992) On magnetotelluric source effects caused by an auroral electrojet system. Radio Sci 27(04): 463–468.

Sato S (2020a) Altitude effects of localized source currents on magnetotelluric responses. Earth Planets Space 72, doi.org/10.1186/s40623-020-01200-7.

Sato S (2020b) Effects of loop source current on magnetotelluric responses of spherical Earth (in Japanese). Butsuri-tansa (Geophys Explor) 73, 168–176.

Sato S, Goto T, Koike K (2020) Spatial gradients of geomagnetic temporal variations causing the instability of interstation transfer functions. Earth Planets Space 72, doi.org/10.1186/s40623-020-01231-0.

Schmucker U (1970) Anomalies of geomagnetic variations in the southwestern United States. Berkeley: Bull Scripps Inst Oceanogr.

Viljanen A (2012) Description of the magnetospheric/ionospheric sources. In: Chave AD, Jones AG

(ed) The Magnetotelluric Method. Cambridge University Press, Cambridge.

Wang H, Egbert GD, Yao Y, Cheng J (2020) Array analysis of magnetic and electric field observatories in China: estimation of magnetotelluric impedances at very long periods. Geophys J Int 222(1): 305–326.

Yamazaki Y, Maute A (2017) Sq and EEJ–A review on the daily variation of the geomagnetic field caused by ionospheric dynamo currents. Space Sci Rev 206: 299–405.